\documentclass[11pt]{article}
\usepackage{amssymb,amsmath,amsfonts}
\usepackage{graphicx}
\usepackage{graphics}
\usepackage{eepic,epsfig}

\begin{document}

\title{Mean field squared and energy-momentum tensor for \\
the hyperbolic vacuum in dS spacetime}
\author{A. A. Saharian$^{1}$,\, T. A. Petrosyan$^{1}$,\, V. S. Torosyan$^{2}$
\\
\\
\textit{$^1$ Department of Physics, Yerevan State University,}\\
\textit{1 Alex Manoogian Street, 0025 Yerevan, Armenia} \vspace{0.3cm}\\
\textit{$^{2}$ Institute of Mechanics NAS RA,}\\
\textit{24b Marshall Baghramian Ave., 0019 Yerevan, Armenia}}
\maketitle

\begin{abstract}
We evaluate the Hadamard function and the vacuum expectation values (VEVs)
of the field squared and energy-momentum tensor for a massless conformally
coupled scalar field in $(D+1)$-dimensional de Sitter (dS) spacetime
foliated by spatial sections of negative constant curvature. It is assumed
that the field is prepared in the hyperbolic vacuum state. An integral
representation for the difference of the Hadamard functions corresponding to
the hyperbolic and Bunch-Davies vacua is provided that is well adapted for
the evaluation of the expectation values in the coincidence limit. It is
shown that the Bunch-Davies state is interpreted as thermal with respect to
the hyperbolic vacuum. An expression for the corresponding density of states
is provided. The relations obtained for the difference in the VEVs for the
Bunch-Davies and hyperbolic vacua are compared with the corresponding
relations for the Fulling-Rindler and Minkowski vacua in flat spacetime. The
similarity between those relations is explained by the conformal connection
of dS spacetime with hyperbolic foliation and Rindler spacetime. As a
limiting case, the VEVs for the conformal vacuum in the Milne universe are
discussed.
\end{abstract}

Keywords: Vacuum polarization, de Sitter spacetime, hyperbolic vacuum

\bigskip

\section{Introduction}

Canonical quantization of fields is carried out by defining complete set of
modes end expanding the field operator in terms of those modes. The creation
and annihilation operators are defined as the coefficients in the expansion.
They allow one to build the Fock space of states starting from the
definition of the vacuum as the state of quantum field nullified by the
annihilation operator. As seen from this construction, the notion of the
vacuum depends on the choice of the normal modes used in the expansion of
the field operator (see \cite{Birr82} for a general discussion). Different
sets of modes, in general, lead to different vacuum states. Well-known
examples of two inequivalent vacua in flat spacetime are the Minkowski
vacuum and the Fulling-Rindler vacuum. They correspond to quantization of
fields in terms of the modes for inertial and uniformly accelerated
observers, respectively. Other examples are the Hartle-Hawking, Boulware and
Unruh vacua for Schwarzschild black hole geometry.

Because of high symmetry and cosmological applications, the de Sitter (dS)
spacetime is among the most important gravitational backgrounds in quantum
field theory. In that geometry there exists a one-parameter family of
dS-invariant vacuum states (known as $\alpha $-vacua) \cite{Alle85,Mott85}.
Among these maximally symmetric states the Bunch-Davies (Euclidean) vacuum
\cite{Bunc78a} is distinguished by the property that in the flat spacetime
limit it is reduced to the Minkowski vacuum. Its black hole counterpart is
the Hartle-Hawking state. The Bunch-Davies (BD) vacuum is naturally realized
in the quantization based on the mode functions being the solutions of
classical field equations in global and planar (inflationary) coordinates of
dS spacetime (for different coordinate systems in dS spacetime, see, for
example \cite{Grif09}). Another coordinate system frequently used in
considerations of thermal aspects of dS spacetime corresponds to static
coordinates. The vacuum state realized by the normal modes in those
coordinates is referred to as static vacuum. In the flat spacetime limit it
reduces to the Fulling-Rindler vacuum and its black hole counterpart is the
Boulware vacuum. The vacuum state in dS spacetime corresponding to the Unruh
vacuum in black hole geometry has been recently considered in \cite{Aals19}
(Unruh-de Sitter state).

Here we will consider the local characteristics of another vacuum state in
dS spacetime referred to as hyperbolic (H-) vacuum \cite{Pfau82}. Its flat
spacetime counterpart corresponds to the conformal vacuum in the Milne patch
of Minkowski spacetime. The H-vacuum is a quantum state that is nullified by
the annihilation operator defined in terms of the normal modes being the
solutions of the field equation in the coordinate system corresponding to
the foliation of dS spacetime by spatial sections having constant negative
curvature ($k=-1$ Robertson-Walker coordinatization of dS spacetime). This
coordinatization has been employed in investigations of open inflationary
models of the early universe (see \cite{Gott82}-\cite{Mats20}) and in
discussions of the entanglement entropy in dS spacetime \cite{Mald13}-\cite%
{Bhat19}. The hyperbolic coordinates are well-adapted for studies of long
range quantum correlations between two causally disconnected regions
separated by a finite region. Recent interest to the open inflation scenario
is related to its natural realization in the context of the string landscape
through the bubble nucleation. In the thin wall approximation, the influence
of the exterior geometry on the dynamics of quantum fluctuations for scalar
fields is reduced to the imposition of Robin type boundary condition (see
\cite{Milt12,Bell14} for the dS geometry described in inflationary and
static coordinates). The corresponding Casimir densities induced by a
spherical boundary in the hyperbolic vacuum of dS spacetime have been
recently discussed in \cite{Saha21}. The effect of a delta-functional wall
on the entanglement entropy between two causally disconnected regions in dS
spacetime with negative curvature spatial foliation is studied in \cite%
{Albr18}.

The paper is organized as follows. In the next section we present the normal
modes and Hadamard functions for the H- and BD vacua for a massive scalar
field with general curvature coupling parameter. The corresponding
expressions are essentially simplified for a conformally coupled massless
field and they are given in section \ref{sec:HFconf}. The VEVs of the field
squared and energy-momentum tensor are investigated in section \ref{sec:VEV}%
. The main results are summarized in section \ref{sec:Conc}.

\section{Hadamard functions in hyperbolic and Bunch-Davies vacua}

\label{sec:Had}

The $(D+1)$-dimensional de Sitter spacetime can be visualized as a
hyperboloid in $(D+2)$-dimensional Minkowski spacetime (see, for instance,
\cite{Grif09}). The entire hyperboloid is covered by the global coordinates.
The corresponding spatial sections are $D$-dimensional spheres. In
cosmological applications, in particular, in models of inflation, the most
popular coordinate system corresponds to the so-called inflationary or
planar coordinates. These coordinates cover a half of the hyperboloid and
the spacetime is foliated by flat spatial sections. Introducing spherical
spatial coordinates $(r_{\mathrm{I}},\vartheta ,\phi )$, with $\vartheta
=(\theta _{1},\ldots \theta _{n})$, $n=D-2$, the line element is presented
in the form%
\begin{equation}
ds^{2}=dt_{\mathrm{I}}^{2}-e^{2t_{\mathrm{I}}/\alpha }\left( dr_{\mathrm{I}%
}^{2}+r_{\mathrm{I}}^{2}d\Omega _{D-1}^{2}\right) ,  \label{dsI}
\end{equation}%
where $-\infty <t_{\mathrm{I}}<+\infty $, $0\leq r_{\mathrm{I}}<\infty $, $%
d\Omega _{D-1}^{2}$ is the line element on a unit sphere $S^{D-1}$, and for
the angular coordinates one has $0\leqslant \theta _{k}\leqslant \pi $, $%
k=1,2,\ldots ,n$, $0\leqslant \phi \leqslant 2\pi $. The parameter $\alpha $
determines the curvature radius of the spacetime and is related to positive
cosmological constant $\Lambda $ by the formula $\Lambda =D(D-1)/(2\alpha
^{2})$.

The foliation of dS spacetime with negative curvature spatial sections is
realized by hyperbolic coordinates $(t,r,\vartheta ,\phi )$ with $0\leq
t,r<\infty $. In these coordinates the line element reads%
\begin{equation}
ds^{2}=dt^{2}-\alpha ^{2}\sinh ^{2}\left( t/\alpha \right) (dr^{2}+\sinh
^{2}rd\Omega _{D-1}^{2}).  \label{dsH}
\end{equation}%
The transformation between the inflationary and hyperbolic coordinates is
given by
\begin{eqnarray}
t_{\mathrm{I}} &=&\alpha \ln \left[ \cosh (t/\alpha )+\sinh (t/\alpha )\cosh
r\right] ,  \notag \\
r_{\mathrm{I}} &=&\alpha e^{-t_{\mathrm{I}}/\alpha }\sinh (t/\alpha )\sinh r.
\label{Infhyp}
\end{eqnarray}%
The discussion of different regions of the Penrose diagram covered by
hyperbolic coordinates can be found in \cite{Saha21,Sasa95}. Passing to a
new radial coordinate $r_{\mathrm{FRW}}=\sinh r$, the line element (\ref{dsH}%
) is written in the form corresponding to open Friedmann-Robertson-Walker
cosmological models with the scale factor $\alpha \sinh \left( t/\alpha
\right) $. In terms of the hyperbolic coordinates, the geodesic distance $%
d(x,x^{\prime })$ between two spacetime points $x=(t,r,\vartheta ,\phi )$
and $x^{\prime }=(t^{\prime },r^{\prime },\vartheta ^{\prime },\phi ^{\prime
})$ is determined by the dS invariant function
\begin{equation}
u(x,x^{\prime })=\cosh (t/\alpha )\cosh (t^{\prime }/\alpha )-\sinh
(t/\alpha )\sinh (t^{\prime }/\alpha )w,  \label{Gd}
\end{equation}%
where%
\begin{equation}
w=w(r,r^{\prime },\theta )=\cosh r\cosh r^{\prime }-\sinh r\sinh r^{\prime
}\cos \theta ,  \label{ub}
\end{equation}%
and $\theta $ is the angle between the directions determined by $(\vartheta
,\phi )$ and $(\vartheta ^{\prime },\phi ^{\prime })$. Note that $w\geq 1$.
For $u(x,x^{\prime })>1$ the geodesic distance is given by the relation $%
\cosh \left[ d(x,x^{\prime })/\alpha \right] =u(x,x^{\prime })$. Otherwise
the hyperbolic function in the left-hand side should be replaced by the
cosine function.

We consider a quantum scalar field $\varphi (x)$ with curvature coupling
parameter $\xi $ in background of dS spacetime. The dynamics of the field is
governed by the equation%
\begin{equation}
\left( \Box +m^{2}+\xi R\right) \varphi =0,  \label{Feq}
\end{equation}%
where $\Box $ is the d'Alembert operator and for the Ricci scalar one has $%
R=D\left( D+1\right) /\alpha ^{2}$. The vacuum state depends on the complete
set of mode functions used in the expansion of the field operator. The mode
functions realizing the H-vacuum are given by the expression (see \cite%
{Dimi15} for the case $D=3$ and \cite{Saha21} in general number of spatial
dimensions)%
\begin{equation}
\varphi _{\sigma }\left( x\right) =c_{\sigma }\frac{P_{\nu -1/2}^{iz}\left(
\cosh (t/\alpha )\right) }{\sinh ^{(D-1)/2}(t/\alpha )}\frac{%
P_{iz-1/2}^{1-D/2-l}\left( \cosh r\right) }{\sinh ^{D/2-1}r}Y\left(
m_{p};\vartheta ,\phi \right) ,  \label{phi}
\end{equation}%
with the normalization coefficient determined from
\begin{equation}
\left\vert c_{\sigma }\right\vert ^{2}=\frac{z\left\vert \Gamma \left( l+%
\frac{D-1}{2}+iz\right) \right\vert ^{2}}{2N\left( m_{p}\right) \alpha ^{D-1}%
}.  \label{C}
\end{equation}%
In (\ref{phi}), $P_{\rho }^{\gamma }(x)$ is the associated Legendre function
of the first kind, $Y(m_{p};\vartheta ,\phi )$ are hyperspherical harmonics
and we have defined%
\begin{equation}
\nu =\sqrt{D^{2}/4-\xi D\left( D+1\right) -m^{2}\alpha ^{2}}.  \label{nun}
\end{equation}%
The modes are specified by the set of quantum numbers $\sigma =\left(
z,m_{p}\right) $, where $0\leq z<\infty $ and for the quantum numbers
related to the angular coordinates one has $m_{p}=(l,m_{1},\ldots ,m_{n})$
with $l=0,1,2,\ldots $. The integers $m_{1},m_{2},\ldots ,m_{n}$ obey the
relations $-m_{n-1}\leqslant m_{n}\leqslant m_{n-1}$ and $0\leqslant
m_{n-1}\leqslant m_{n-2}\leqslant \cdots \leqslant m_{1}\leqslant l$.

Given the mode functions (\ref{phi}), we can evaluate the Hadamard function
for the H-vacuum by using the mode-sum formula
\begin{equation}
G(x,x^{\prime })=\sum_{\sigma }\left[ \varphi _{\sigma }\left( x\right)
\varphi _{\sigma }^{\ast }\left( x^{\prime }\right) +\varphi _{\sigma
}\left( x^{\prime }\right) \varphi _{\sigma }^{\ast }\left( x\right) \right]
.  \label{WF}
\end{equation}%
In \cite{Saha21} the following representation has been derived:
\begin{eqnarray}
G\left( x,x^{\prime }\right) &=&\frac{\alpha ^{1-D}}{2(2\pi )^{D/2}}%
\int_{0}^{\infty }dz\,z\left\vert \Gamma \left( \frac{D-1}{2}+iz\right)
\right\vert ^{2}  \notag \\
&&\times \frac{\sum_{j=+,-}P_{\nu -1/2}^{jiz}\left( \cosh (t/\alpha )\right)
P_{\nu -1/2}^{-jiz}\left( \cosh (t^{\prime }/\alpha )\right) }{\left[ \sinh
(t/\alpha )\sinh (t^{\prime }/\alpha )\right] ^{(D-1)/2}}\frac{%
P_{iz-1/2}^{1-D/2}\left( w\right) }{\left( w^{2}-1\right) ^{(D-2)/4}},
\label{wfdsh}
\end{eqnarray}%
where $w$ is defined by (\ref{ub}).

The vacuum state for a scalar field most frequently used in the inflationary
coordinates is the BD vacuum. It is realized by the mode functions (for the
corresponding mode functions in hyperbolic foliation see \cite{Sasa95})%
\begin{equation}
\varphi _{\sigma }(x)=\frac{c_{\mathrm{I}\sigma }\eta _{\mathrm{I}}^{D/2}}{%
r_{\mathrm{I}}^{D/2-1}}H_{\nu }^{(1)}(\lambda |\eta _{\mathrm{I}}|)J_{\mu
}(\lambda r_{\mathrm{I}})Y(m_{p};\vartheta ,\phi ),  \label{eigfunc}
\end{equation}%
where $H_{\nu }^{(1)}(y)$ and $J_{\mu }(y)$ are the Hankel and Bessel
functions, respectively, $\eta _{\mathrm{I}}=-\alpha e^{-t_{\mathrm{I}%
}/\alpha }$, $-\infty <\eta _{\mathrm{I}}<0$, is the inflationary conformal
time coordinate. The normalization coefficient is given by
\begin{equation}
\left\vert c_{\mathrm{I}\sigma }\right\vert ^{2}=\frac{\pi e^{i(\nu -\nu
^{\ast })\pi /2}}{4N(m_{p})\alpha ^{D-1}}.  \label{CI}
\end{equation}%
The expression for the corresponding Hadamard function reads (for the
corresponding Wightman function see \cite{Cand75})%
\begin{equation}
G_{\mathrm{BD}}\left( x,x^{\prime }\right) =\frac{\Gamma \left( D/2+\nu
\right) \Gamma \left( D/2-\nu \right) }{\left( 2\pi \right) ^{\left(
D+1\right) /2}\alpha ^{D-1}}\frac{P_{\nu -1/2}^{\left( 1-D\right) /2}\left(
-u(x,x^{\prime })\right) }{|u^{2}(x,x^{\prime })-1|^{\left( D-1\right) /4}}.
\label{wfdsi}
\end{equation}%
In terms of the inflationary coordinates the function $u(x,x^{\prime })$ is
written as%
\begin{equation}
u(x,x^{\prime })=\frac{(\Delta \eta _{\mathrm{I}})^{2}-|\Delta \mathbf{r}_{%
\mathrm{I}}|^{2}}{2\eta _{\mathrm{I}}\eta _{\mathrm{I}}^{\prime }}+1,
\label{uI}
\end{equation}%
where $\Delta \eta _{\mathrm{I}}=\eta _{\mathrm{I}}^{\prime }-\eta _{\mathrm{%
I}}$ and $|\Delta \mathbf{r}_{\mathrm{I}}|^{2}=r_{\mathrm{I}}^{2}+r_{\mathrm{%
I}}^{\prime 2}-2r_{\mathrm{I}}r_{\mathrm{I}}^{\prime }\cos \theta $. Note
that passing to the time coordinate $\eta _{\mathrm{I}}$ the line element (%
\ref{dsI}) is written in manifestly conformally flat form as
\begin{equation}
ds^{2}=\left( \alpha /\eta _{\mathrm{I}}\right) ^{2}\left( d\eta _{\mathrm{I}%
}^{2}-dr_{\mathrm{I}}^{2}-r_{\mathrm{I}}^{2}d\Omega _{D-1}^{2}\right) .
\label{ds2Icf}
\end{equation}%
The Hadamard function for the BD vacuum depends on the spacetime points
through the geodesic distance and that state is maximally symmetric. As it
has been mentioned above, in dS spacetime there is a one-parameter family of
maximally symmetric vacuum states called as $\alpha $-vacua.

\section{Hadamard function for a conformally coupled massless field}

\label{sec:HFconf}

The expressions for the Hadamard functions are simplified for a conformally
coupled massless field with $\xi =(D-1)/(4D)$ and $m=0$. In this case $\nu
=1/2$ and for the functions in the integrand of (\ref{wfdsh}) we use
\begin{equation}
P_{0}^{\pm iz}\left( \cosh \left( t/\alpha \right) \right) =\frac{e^{\mp
iz\eta /\alpha }}{\Gamma \left( 1\mp iz\right) },  \label{P0}
\end{equation}%
where $\eta $, $-\infty <\eta \leq 0$, is the conformal time coordinate in
the hyperbolic foliation. One has the following relation $\cosh \left(
t/\alpha \right) =-\coth \left( \eta /\alpha \right) $. From (\ref{P0}) we
see that time dependence of the normal modes appears in the form $e^{-iz\eta
/\alpha }$ and for the energy $E$ related to the conformal time coordinate
one gets $E=z/\alpha $. The Hadamard function for the H-vacuum takes the form%
\begin{eqnarray}
G\left( x,x^{\prime }\right) &=&\frac{2\left[ \sinh (\eta /\alpha )\sinh
(\eta ^{\prime }/\alpha )\right] ^{(D-1)/2}}{(2\pi )^{D/2+1}\alpha
^{D-1}\left( w^{2}-1\right) ^{(D-2)/4}}\int_{0}^{\infty }dz\,\sinh \left(
\pi z\right)  \notag \\
&&\times \cos \left( z\Delta \eta /\alpha \right) \left\vert \Gamma \left(
\frac{D-1}{2}+iz\right) \right\vert ^{2}P_{iz-1/2}^{1-D/2}\left( w\right) ,
\label{Gconf}
\end{eqnarray}%
where $\Delta \eta =\eta ^{\prime }-\eta $.

By using the properties of the associated Legendre function the following
relation can be proved:%
\begin{equation}
\left\vert \Gamma \left( \frac{D-1}{2}+iz\right) \right\vert ^{2}\frac{%
P_{iz-1/2}^{1-D/2}\left( w\right) }{\left( w^{2}-1\right) ^{\frac{D-2}{4}}}%
=\left( -1\right) ^{n}\left\vert \Gamma \left( \frac{D-1}{2}-n+iz\right)
\right\vert ^{2}\partial _{w}^{n}\frac{P_{iz-1/2}^{1-D/2+n}\left( w\right) }{%
\left( w^{2}-1\right) ^{\frac{D-2n-2}{4}}},  \label{RelP}
\end{equation}%
with $n$ being a non-negative integer. This allows to present the Hadamard
function in the form%
\begin{eqnarray}
G\left( x,x^{\prime }\right) &=&2\left( -1\right) ^{n}\frac{\left[ \sinh
(\eta /\alpha )\sinh (\eta ^{\prime }/\alpha )\right] ^{\frac{D-1}{2}}}{%
(2\pi )^{D/2+1}\alpha ^{D-1}}\partial _{w}^{n}\int_{0}^{\infty }dz\,\sinh
\left( \pi z\right)  \notag \\
&&\times \cos \left( z\Delta \eta /\alpha \right) \left\vert \Gamma \left(
\frac{D-1}{2}-n+iz\right) \right\vert ^{2}\frac{P_{iz-1/2}^{1-D/2+n}\left(
w\right) }{\left( w^{2}-1\right) ^{\frac{D-2n-2}{4}}}.  \label{Gconf2}
\end{eqnarray}%
This expression can be further simplified and we will consider the odd and
even values for the spatial dimension $D$ separately.

In terms of the conformal time, by taking into account that $\sinh \left(
t/\alpha \right) =-1/\sinh \left( \eta /\alpha \right) $, the line element (%
\ref{dsH}) reads%
\begin{equation}
ds^{2}=\sinh ^{-2}\left( \eta /\alpha \right) \left[ d\eta ^{2}-\alpha
^{2}\left( dr^{2}+\sinh ^{2}rd\Omega _{D-1}^{2}\right) \right] .
\label{ds2c}
\end{equation}%
This shows that the geometry we consider is conformally related to a static
spacetime foliated by constant negative curvature spatial sections with the
conformal factor $\Omega _{\mathrm{st}}^{2}\left( \eta \right) =\sinh
^{-2}\left( \eta /\alpha \right) $. The expression for the corresponding
Hadamard function, denoted here by $G_{\mathrm{st}}\left( x,x^{\prime
}\right) $, is directly obtained from (\ref{Gconf}) and from the relation
\begin{equation}
G\left( x,x^{\prime }\right) =\left[ \Omega _{\mathrm{st}}\left( \eta
\right) \Omega _{\mathrm{st}}\left( \eta ^{\prime }\right) \right] ^{\frac{%
1-D}{2}}G_{\mathrm{st}}\left( x,x^{\prime }\right) .  \label{GrelSt}
\end{equation}%
An alternative expression for $G_{\mathrm{st}}\left( x,x^{\prime }\right) $
is obtained from (\ref{Gconf2}).

\subsection{Odd $D$}

For odd values of $D$, we take $n=(D-3)/2$ and in the right-hand side of (%
\ref{RelP}) we use the relation%
\begin{equation}
P_{iz-1/2}^{-1/2}\left( w\right) =\sqrt{\frac{2}{\pi }}\frac{\sin \left(
z\zeta \right) }{z\sqrt{\sinh \zeta }},  \label{Pm12}
\end{equation}%
where $\zeta $ is defined in accordance with%
\begin{equation}
w=\cosh \zeta .  \label{zeta}
\end{equation}%
This leads to the expression%
\begin{equation}
G\left( x,x^{\prime }\right) =\frac{\left[ \sinh (\eta /\alpha )\sinh (\eta
^{\prime }/\alpha )\right] ^{\frac{D-1}{2}}}{(-2\pi )^{\frac{D+1}{2}}\alpha
^{D-1}}\,\left( \frac{\partial _{\zeta }}{\sinh \zeta }\right) ^{\frac{D-3}{2%
}}\frac{2\zeta /\sinh \zeta }{\zeta ^{2}-\left( \Delta \eta \right)
^{2}/\alpha ^{2}},  \label{GconfO}
\end{equation}%
for the Hadamard function. The corresponding formula for the static open
universe is obtained from (\ref{GrelSt}). In the special case $D=3$ and $%
\theta =0$ it is in agreement with the expression for the Wightman function
given in \cite{Bunc78}. The corresponding conformally transformed Hadamard
function has been used in \cite{Pfau82} for the evaluation of the VEV for
the energy-momentum tensor in the H-vacuum of dS spacetime.

For the BD vacuum, by using the relation%
\begin{equation}
P_{0}^{-\beta }\left( -u\right) =\frac{1}{\Gamma \left( 1+\beta \right) }%
\left( \frac{u+1}{u-1}\right) ^{\beta /2},  \label{P0b}
\end{equation}%
from (\ref{wfdsi}) we get%
\begin{equation}
G_{\mathrm{BD}}\left( x,x^{\prime }\right) =\frac{\alpha ^{1-D}\Gamma \left(
(D-1)/2\right) }{\left( 2\pi \right) ^{\frac{D+1}{2}}\left[ 1-u(x,x^{\prime
})\right] ^{\frac{D-1}{2}}}.  \label{GBDconf}
\end{equation}%
In the case of odd $D$ this expression can be rewritten as%
\begin{equation}
G_{\mathrm{BD}}\left( x,x^{\prime }\right) =\frac{\alpha ^{1-D}}{\left( 2\pi
\right) ^{\frac{D+1}{2}}}\partial _{u}^{\frac{D-3}{2}}\frac{1}{%
1-u(x,x^{\prime })}.  \label{GBDconf2}
\end{equation}%
For the further transformation of this expression we note that%
\begin{equation}
\partial _{u}=-\sinh (\eta /\alpha )\sinh (\eta ^{\prime }/\alpha )\frac{%
\partial _{\zeta }}{\sinh \zeta },  \label{du}
\end{equation}%
and
\begin{equation}
1-u(x,x^{\prime })=\frac{\cosh \zeta -\cosh \left( \Delta \eta /\alpha
\right) }{\sinh (\eta /\alpha )\sinh (\eta ^{\prime }/\alpha )}.  \label{up1}
\end{equation}%
The Hadamard function is expressed as
\begin{equation}
G_{\mathrm{BD}}\left( x,x^{\prime }\right) =\frac{\left[ \sinh (\eta /\alpha
)\sinh (\eta ^{\prime }/\alpha )\right] ^{\frac{D-1}{2}}}{\left( -2\pi
\right) ^{\frac{D+1}{2}}\alpha ^{D-1}}\left( \frac{\partial _{\zeta }}{\sinh
\zeta }\right) ^{\frac{D-3}{2}}\frac{1}{\cosh \zeta -\cosh \left( \Delta
\eta /\alpha \right) }.  \label{GBDconf3}
\end{equation}%
This representation is well adapted for the evaluation of the difference in
the VEVs for the H- and BD vacua.

The BD vacuum is conformally related to the Minkowski vacuum in flat
spacetime and for a massless conformally coupled scalar field the relation $%
G_{\mathrm{BD}}\left( x,x^{\prime }\right) =(\eta _{\mathrm{I}}\eta _{%
\mathrm{I}}^{\prime }/\alpha ^{2})^{(D-1)/2}G_{\mathrm{M}}\left( x,x^{\prime
}\right) $ is expected between the corresponding Hadamard functions. This
can be seen by using the standard expression
\begin{equation}
G_{\mathrm{M}}\left( x,x^{\prime }\right) =\frac{\Gamma \left(
(D-1)/2\right) }{2\pi ^{\left( D+1\right) /2}}\left( |\Delta \mathbf{r}_{%
\mathrm{I}}|^{2}-\Delta \eta _{\mathrm{I}}^{2}\right) ^{\frac{1-D}{2}},
\label{GM1}
\end{equation}%
and passing to the hyperbolic coordinates. That leads to the following
representation
\begin{equation}
G_{\mathrm{M}}\left( x,x^{\prime }\right) =\frac{\Gamma \left(
(D-1)/2\right) }{\left( 2\pi \right) ^{\frac{D+1}{2}}\left( \eta _{\mathrm{I}%
}\eta _{\mathrm{I}}^{\prime }\right) ^{\frac{D-1}{2}}}\left[ \frac{\sinh
(\eta /\alpha )\sinh (\eta ^{\prime }/\alpha )}{\cosh \zeta -\cosh \left(
\Delta \eta /\alpha \right) }\right] ^{\frac{D-1}{2}}.  \label{GM2}
\end{equation}%
Multiplying this by $(\eta _{\mathrm{I}}\eta _{\mathrm{I}}^{\prime }/\alpha
^{2})^{(D-1)/2}$ we get the function (\ref{GBDconf3}).

The difference of the Hadamard functions, that determines the difference in
the local VEVs, is presented as%
\begin{eqnarray}
&& G\left( x,x^{\prime }\right) -G_{\mathrm{BD}}\left( x,x^{\prime }\right) =%
\frac{\left[ \sinh (\eta /\alpha )\sinh (\eta ^{\prime }/\alpha )\right] ^{%
\frac{D-1}{2}}}{(-2\pi )^{\frac{D+1}{2}}\alpha ^{D-1}}\,  \notag \\
&& \qquad \times \left( \frac{\partial _{\zeta }}{\sinh \zeta }\right) ^{%
\frac{D-3}{2}}\left[ \frac{2\zeta /\sinh \zeta }{\zeta ^{2}-\left( \Delta
\eta \right) ^{2}/\alpha ^{2}}-\frac{1}{\cosh \zeta -\cosh \left( \Delta
\eta /\alpha \right) }\right] .  \label{Gdif}
\end{eqnarray}%
Note that in this formula the part
\begin{equation}
G_{\mathrm{st}}\left( x,x^{\prime }\right) =\frac{\alpha ^{1-D}}{(-2\pi )^{%
\frac{D+1}{2}}}\left( \frac{\partial _{\zeta }}{\sinh \zeta }\right) ^{\frac{%
D-3}{2}}\frac{2\zeta /\sinh \zeta }{\zeta ^{2}-\left( \Delta \eta \right)
^{2}/\alpha ^{2}},  \label{Gst}
\end{equation}%
with $x=(\eta ,r,\vartheta ,\phi )$, is the Hadamard function for a scalar
field in static hyperbolic universe with the line element given by the
expression in the square brackets of (\ref{ds2c}). By using the result
\begin{equation}
\sum_{n=-\infty }^{\infty }\frac{a}{n^{2}+a^{2}}=\pi \coth (\pi a),
\label{Ser}
\end{equation}%
the following relation is proved:%
\begin{equation}
\sum_{n=-\infty }^{+\infty }G_{\mathrm{st}}\left( x_{n\beta },x^{\prime
}\right) =\frac{\alpha ^{1-D}}{(-2\pi )^{\frac{D+1}{2}}}\left( \frac{%
\partial _{\zeta }}{\sinh \zeta }\right) ^{\frac{D-3}{2}}\frac{y\sinh
(y\zeta )/\sinh \zeta }{\cosh (y\zeta )-\cosh (y\Delta \eta /\alpha )},
\label{Gstbet}
\end{equation}%
where $x_{n\beta }=(\eta +in\beta ,r,\vartheta ,\phi )$ and $y=2\pi \alpha
/\beta $. Taking $\beta =2\pi \alpha $, we see that the state corresponding
to the part with the second term in the square brackets of (\ref{Gdif}) is
interpreted as a thermal state with temperature $1/(2\pi \alpha )$. This
result for $D=3$ has been discussed in \cite{Pfau82}.

\subsection{Even $D$}

For even values of $D$, taking $n=D/2-1$, the expression of the Hadamard
function for the H-vacuum reads%
\begin{eqnarray}
G\left( x,x^{\prime }\right) &=&-\frac{\alpha ^{1-D}}{(-2\pi )^{D/2}}\left[
\sinh (\eta /\alpha )\sinh (\eta ^{\prime }/\alpha )\right] ^{\frac{D-1}{2}}
\notag \\
&&\times \partial _{w}^{D/2-1}\int_{0}^{\infty }dz\,\tanh \left( \pi
z\right) \cos \left( z\Delta \eta /\alpha \right) P_{iz-1/2}\left( w\right) .
\label{GconfEv}
\end{eqnarray}%
For the BD vacuum one has the expression (\ref{GBDconf}). By taking into
account the relation
\begin{equation}
\frac{1}{y^{\frac{D-1}{2}}}=\frac{\left( -1\right) ^{D/2-1}\pi ^{1/2}}{%
\Gamma ((D-1)/2)}\partial _{y}^{D/2-1}\frac{1}{\sqrt{y}}  \label{Dif}
\end{equation}%
and (\ref{up1}), it can be rewritten in the form
\begin{equation}
G_{\mathrm{BD}}\left( x,x^{\prime }\right) =-\frac{\left[ \sinh (\eta
/\alpha )\sinh (\eta ^{\prime }/\alpha )\right] ^{\frac{D-1}{2}}}{\sqrt{2}%
\left( -2\pi \right) ^{\frac{D}{2}}\alpha ^{D-1}}\partial _{w}^{D/2-1}\frac{1%
}{\sqrt{w-\cosh \left( \Delta \eta /\alpha \right) }}.  \label{Rel2}
\end{equation}%
Next we use the relation \cite{Prud2}%
\begin{equation}
\int_{0}^{\infty }dz\,\cos \left( z\Delta \eta /\alpha \right)
P_{iz-1/2}\left( w\right) =\frac{1}{\sqrt{2}\left[ w-\cosh \left( \Delta
\eta /\alpha \right) \right] ^{1/2}},  \label{IntForm}
\end{equation}%
to present the Hadamard function in the form%
\begin{eqnarray}
G_{\mathrm{BD}}\left( x,x^{\prime }\right) &=&-\frac{\alpha ^{1-D}}{\left(
-2\pi \right) ^{D/2}}\left[ \sinh (\eta /\alpha )\sinh (\eta ^{\prime
}/\alpha )\right] ^{\frac{D-1}{2}}  \notag \\
&&\times \partial _{w}^{D/2-1}\int_{0}^{\infty }dz\,\cos \left( z\Delta \eta
/\alpha \right) P_{iz-1/2}\left( w\right) .  \label{GBDconf4}
\end{eqnarray}%
For the difference of the Hadamard functions we get%
\begin{eqnarray}
G\left( x,x^{\prime }\right) -G_{\mathrm{BD}}\left( x,x^{\prime }\right) &=&%
\frac{2\alpha ^{1-D}}{(-2\pi )^{\frac{D}{2}}}\left[ \sinh (\eta /\alpha
)\sinh (\eta ^{\prime }/\alpha )\right] ^{\frac{D-1}{2}}  \notag \\
&&\times \int_{0}^{\infty }dz\,\frac{\cos \left( z\Delta \eta /\alpha
\right) }{e^{2\pi z}+1}\frac{P_{iz-1/2}^{D/2-1}\left( w\right) }{(w^{2}-1)^{%
\frac{D-2}{4}}}.  \label{Gdif2}
\end{eqnarray}%
where we have introduced the associated Legendre function of the first kind
in accordance with $\partial _{w}^{D/2-1}P_{iz-1/2}\left( w\right)
=(w^{2}-1)^{(2-D)/4}P_{iz-1/2}^{D/2-1}\left( w\right) $.

In the discussion above we have considered the difference in the Hadamard
functions for the H- and BD vacua. Similar expressions are obtained for the
differences of other two-point functions. For example, the difference in the
Wightman functions for odd and even values of spatial dimension is given by
the right-hand sides of (\ref{Gdif}) and (\ref{Gdif2}) with an additional
coefficient 1/2. The corresponding formulas can be used for the
investigation of the response of particle detectors.

\section{Vacuum expectation values}

\label{sec:VEV}

In this section we evaluate the VEVs of the field squared and of the
energy-momentum tensor for the H-vacuum by using the formulas (\ref{Gdif})
and (\ref{Gdif2}).

\subsection{Mean field squared}

We start the investigation for local VEVs from the mean field squared.
Having the difference between the Hadamard functions, it can be evaluated by
making use of the formula%
\begin{equation}
\left\langle \varphi ^{2}\right\rangle =\left\langle \varphi
^{2}\right\rangle _{\mathrm{BD}}+\frac{1}{2}\underset{x^{\prime }\rightarrow
x}{\lim }\left[ G\left( x,x^{\prime }\right) -G_{\mathrm{BD}}\left(
x,x^{\prime }\right) \right] .  \label{phi2}
\end{equation}%
The important thing to be mentioned here is that the last term is finite and
the renormalization of the divergences is reduced to the one for $%
\left\langle \varphi ^{2}\right\rangle _{\mathrm{BD}}$. The latter procedure
is widely discussed in the literature. Here we note that, because of the
maximal symmetry of the BD vacuum, the VEV $\left\langle \varphi
^{2}\right\rangle _{\mathrm{BD}}$ does not depend on the spacetime
coordinates.

We start from the case of even $D$. By taking into account that for $%
w\rightarrow 1+$ one has \cite{Nist10}
\begin{equation}
P_{iz-1/2}^{\mu }\left( w\right) \approx \frac{\Gamma \left( iz+1/2+\mu
\right) }{\Gamma (\mu +1)\Gamma \left( iz+1/2-\mu \right) }\left( \frac{w-1}{%
2}\right) ^{\mu /2},  \label{P1}
\end{equation}%
we get%
\begin{equation}
\lim_{w\rightarrow 1}\frac{P_{iz-1/2}^{D/2-1}\left( w\right) }{(w^{2}-1)^{%
\frac{D-2}{4}}}=\frac{2^{1-D/2}\Gamma \left( iz+(D-1)/2\right) }{\Gamma
(D/2)\Gamma \left( iz-(D-3)/2\right) }.  \label{RelPl}
\end{equation}%
By using the relations for the gamma function, the VEV is presented as%
\begin{equation}
\left\langle \varphi ^{2}\right\rangle =\left\langle \varphi
^{2}\right\rangle _{\mathrm{BD}}-\frac{\left[ \alpha \sinh (t/\alpha )\right]
^{1-D}}{2^{D}\pi ^{D/2+1}\Gamma (D/2)}\int_{0}^{\infty }dz\,e^{-\pi
z}\left\vert \Gamma \left( \left( D-1\right) /2+iz\right) \right\vert ^{2}.
\label{phi2ev}
\end{equation}%
An equivalent expression is given by
\begin{equation}
\left\langle \varphi ^{2}\right\rangle =\left\langle \varphi
^{2}\right\rangle _{\mathrm{BD}}-\frac{2\left[ \alpha \sinh (t/\alpha )%
\right] ^{1-D}}{(4\pi )^{D/2}\Gamma (D/2)}\int_{0}^{\infty }dz\,\frac{z^{D-2}%
}{e^{2\pi z}+1}\prod_{l=0}^{D/2-2}\left[ \left( \frac{l+1/2}{z}\right) ^{2}+1%
\right] ,  \label{phi2ev2}
\end{equation}%
where the integral is expressed in terms of the Riemann zeta function. Note
that the difference $\left\langle \varphi ^{2}\right\rangle -\left\langle
\varphi ^{2}\right\rangle _{\mathrm{BD}}$ is negative.

Now we turn to the case of odd $D$. In the evaluation of the coincidence
limit for (\ref{Gdif}) first we put $\theta =0$ and $\Delta \eta =0$. In
this case one gets $\zeta =\Delta r=r^{\prime }-r$. For the mean field
squared we find%
\begin{equation}
\left\langle \varphi ^{2}\right\rangle =\left\langle \varphi
^{2}\right\rangle _{\mathrm{BD}}+\frac{(-2\pi )^{-\frac{D+1}{2}}}{2\left[
\alpha \sinh (t/\alpha )\right] ^{D-1}}\,\lim_{w\rightarrow 1}\partial _{w}^{%
\frac{D-3}{2}}g(w),  \label{phi2oddn1}
\end{equation}%
with the function%
\begin{equation}
g(w)=\frac{2}{\mathrm{arccosh\,}\left( w\right) \sqrt{w^{2}-1}}-\frac{1}{w-1}%
.  \label{gu}
\end{equation}%
We define the constants $a_{l}$, $l=0,1,2,\ldots $ in accordance with%
\begin{equation}
g(w)=\frac{1}{6}\sum_{l=0}^{\infty }a_{l}\left( w-1\right) ^{l}.
\label{gexp}
\end{equation}%
For the first five coefficients one has%
\begin{equation}
a_{0}=-1,\;a_{1}=\frac{11}{30},\;a_{2}=-\frac{191}{1260},\;a_{3}=\frac{2497}{%
37800},\;a_{4}=-\frac{14797}{498960}.  \label{al2}
\end{equation}

The mean field squared is expressed as%
\begin{equation}
\left\langle \varphi ^{2}\right\rangle =\left\langle \varphi
^{2}\right\rangle _{\mathrm{BD}}-\frac{(2\pi )^{-\frac{D+1}{2}}b_{D}}{12%
\left[ \alpha \sinh (t/\alpha )\right] ^{D-1}}\,.  \label{phi2odd}
\end{equation}%
where
\begin{equation}
b_{D}=\left( -1\right) ^{\frac{D-1}{2}}\Gamma ((D-1)/2)a_{(D-3)/2}.
\label{bD2}
\end{equation}%
Note that the constants $b_{D}$ can be directly expressed as
\begin{equation}
b_{D}=6\left( -1\right) ^{\frac{D-1}{2}}\lim_{w\rightarrow 1}\partial _{w}^{%
\frac{D-3}{2}}g(w).  \label{bD}
\end{equation}%
For the first five coefficients we get%
\begin{equation}
b_{3}=1,\;b_{5}=\frac{11}{30},\;b_{7}=\frac{191}{630},\;b_{9}=\frac{2497}{%
6300},\;b_{11}=\frac{14797}{20790}.  \label{bD1}
\end{equation}%
Similar to the previous case, the difference $\left\langle \varphi
^{2}\right\rangle -\left\langle \varphi ^{2}\right\rangle _{\mathrm{BD}}$ is
negative.

It can be checked that, similar to (\ref{phi2ev2}), the result (\ref{phi2odd}%
) is written in the integral form as%
\begin{equation}
\left\langle \varphi ^{2}\right\rangle =\left\langle \varphi
^{2}\right\rangle _{\mathrm{BD}}-\frac{2\left[ \alpha \sinh (t/\alpha )%
\right] ^{1-D}}{(4\pi )^{D/2}\Gamma (D/2)}\int_{0}^{\infty }dz\,\frac{z^{D-2}%
}{e^{2\pi z}-1}\prod_{l=0}^{(D-3)/2}\left[ \left( \frac{l}{z}\right) ^{2}+1%
\right] .  \label{phi2odd2}
\end{equation}%
Combining the formulas given above, for the mean field squared we write
\begin{equation}
\left\langle \varphi ^{2}\right\rangle =\left\langle \varphi
^{2}\right\rangle _{\mathrm{BD}}-\frac{B_{D}}{\left[ \alpha \sinh (t/\alpha )%
\right] ^{D-1}},  \label{phi22}
\end{equation}%
where the coefficient is given by the expression
\begin{equation}
B_{D}=\frac{2(4\pi )^{-\frac{D}{2}}}{\Gamma (D/2)}\int_{0}^{\infty }dz\,%
\frac{z^{D-2}A_{D}(z)}{e^{2\pi z}+(-1)^{D}}.  \label{BD}
\end{equation}%
Here we have introduced the function%
\begin{equation}
A_{D}(z)=\prod_{l=0}^{l_{m}}\left[ \left( \frac{l+1/2-\{D/2\}}{z}\right)
^{2}+1\right] ,  \label{AD}
\end{equation}%
where $l_{m}=D/2-2+\{D/2\}$ and $\{D/2\}$ stands for the fractional part of $%
D/2$.

The relation (\ref{phi22}) between the VEVs in the H- and BD vacua is
similar to the corresponding relations between the Fulling-Rindler and
Minkowski vacua in flat spacetime. In the Rindler coordinates $(\tau _{%
\mathrm{R}},\chi ,\mathbf{x}_{\mathrm{R}})$, with $\mathbf{x}_{\mathrm{R}%
}=(x_{\mathrm{R}}^{2},\ldots ,x_{\mathrm{R}}^{D})$, the Minkowskian line
element is written as
\begin{equation}
ds_{\mathrm{M}}^{2}=\chi ^{2}d\tau _{\mathrm{R}}^{2}-d\chi ^{2}-d\mathbf{x}_{%
\mathrm{R}}^{2},  \label{ds2Ri}
\end{equation}%
and the mean field squared for a scalar massless field in the
Fulling-Rindler and Minkowski vacua are connected by the relation \cite%
{Saha02}%
\begin{equation}
\left\langle \varphi ^{2}\right\rangle _{\mathrm{FR}}=\left\langle \varphi
^{2}\right\rangle _{\mathrm{M}}-B_{D}\chi ^{1-D},  \label{ph2FR}
\end{equation}%
with the same coefficient $B_{D}$ from (\ref{BD}). The worldline with fixed
spatial coordinates $(\chi ,\mathbf{x}_{\mathrm{R}})$ describes an observer
with constant proper acceleration $1/\chi $.

\subsection{VEV of the energy-momentum tensor}

Another important local characteristic of the vacuum state is the VEV of the
energy-momentum tensor. The difference in the VEVs for the H- and BD vacua, $%
\Delta \left\langle T_{ik}\right\rangle =\left\langle T_{ik}\right\rangle
-\left\langle T_{ik}\right\rangle _{\mathrm{BD}}$, can be evaluated by
making use of the formula
\begin{eqnarray}
\Delta \left\langle T_{ik}\right\rangle &=&\frac{1}{2}\underset{x^{\prime
}\rightarrow x}{\lim }\partial _{i^{\prime }}\partial _{k}\left[ G\left(
x,x^{\prime }\right) -G_{\mathrm{BD}}\left( x,x^{\prime }\right) \right]
\notag \\
&&+\left[ \left( \xi -\frac{1}{4}\right) g_{ik}\nabla _{p}\nabla ^{p}-\xi
\nabla _{i}\nabla _{k}-\xi R_{ik}\right] \left( \left\langle \varphi
^{2}\right\rangle -\left\langle \varphi ^{2}\right\rangle _{\mathrm{BD}%
}\right) ,  \label{TikVev}
\end{eqnarray}%
where $\xi =(D-1)/(4D)$ for a conformally coupled field, $\nabla _{i}$ is
the covariant derivative operator and $R_{ik}=Dg_{ik}/\alpha ^{2}$ is the
Ricci tensor for the dS spacetime. The expression in the right-hand side is
finite and a renormalization is not required. It does not contain
ambiguities that are present in various schemes of renormalization for
separate parts $\left\langle T_{ik}\right\rangle $ and $\left\langle
T_{ik}\right\rangle _{\mathrm{BD}}$ (for discussion of renormalization
ambiguities in the VEV\ of the energy-momentum tensor see \cite%
{More03,Holl05,Deca08} and references therein). From the symmetry of the
problem we expect that the VEV (\ref{TikVev}) is a function of the time
coordinate alone and the vacuum stresses are isotropic $\Delta \left\langle
T_{1}^{1}\right\rangle =\Delta \left\langle T_{2}^{2}\right\rangle =\cdots
=\Delta \left\langle T_{D}^{D}\right\rangle $. The continuity equation $%
\nabla _{k}\left\langle T_{i}^{k}\right\rangle =0$ leads to the relation%
\begin{equation}
\Delta \left\langle T_{1}^{1}\right\rangle =\frac{\partial _{t/\alpha }\left[
\sinh ^{D}(t/\alpha )\Delta \left\langle T_{0}^{0}\right\rangle \right] }{%
D\sinh ^{D-1}(t/\alpha )\cosh \left( t/\alpha \right) },  \label{conteq}
\end{equation}%
between the energy density and the stresses. We consider the case of a
conformally coupled massless field and the tensor (\ref{TikVev}) is
traceless $\Delta \left\langle T_{i}^{i}\right\rangle =0$. This leads to the
relation $\Delta \left\langle T_{0}^{0}\right\rangle =-D\Delta \left\langle
T_{1}^{1}\right\rangle $. Combining this with (\ref{conteq}) we conclude
that the tensor $\left\langle T_{i}^{k}\right\rangle $ has the structure
\begin{equation}
\left\langle T_{i}^{k}\right\rangle =\left\langle T_{i}^{k}\right\rangle _{%
\mathrm{BD}}+C_{D}\frac{\mathrm{diag}\left( 1,-1/D,\cdots ,-1/D\right) }{%
\left[ \alpha \sinh (t/\alpha )\right] ^{D+1}},  \label{Tstruc}
\end{equation}%
where the VEV for the BD vacuum has the form $\left\langle
T_{i}^{k}\right\rangle _{\mathrm{BD}}=C_{D}^{\mathrm{(BD)}}\delta
_{i}^{k}\alpha ^{-D-1}/(D+1)$ with a numerical constant $C_{D}^{\mathrm{(BD)}%
}$. Note that the latter is completely determined by the trace anomaly: $%
C_{D}^{\mathrm{(BD)}}=\alpha ^{D+1}\left\langle T_{i}^{i}\right\rangle _{%
\mathrm{BD}}$. In odd dimensional spacetimes the trace anomaly is absent and
$C_{D}^{\mathrm{(BD)}}=0$. For $D=3$ one has $C_{3}^{\mathrm{(BD)}%
}=1/(240\pi ^{2})$ and the trace anomaly in odd dimensional dS spacetimes
with $D>3$ has been investigated in \cite{Cope86}. In particular, $C_{5}^{%
\mathrm{(BD)}}=-5/(4032\pi ^{3})$ and $C_{7}^{\mathrm{(BD)}}=23/(34560\pi
^{4})$. The sign of the coefficient $C_{D}^{\mathrm{(BD)}}$ is determined by
$(-1)^{(D-3)/2}$. The conformal anomaly in spaces with hyperbolic spatial
sections was considered in \cite{Byts95} (for a recent discussion of
renormalization of the energy-momentum tensor in a general spacetime of
arbitrary dimension see \cite{Deca08}). As seen from (\ref{Tstruc}), for the
evaluation of the difference in the VEVs for the BD and H-vacua it is
sufficient to evaluate one of the components in (\ref{TikVev}).

We will consider the component $\Delta \left\langle T_{11}\right\rangle $.
First of all it can be seen that%
\begin{eqnarray}
\nabla _{p}\nabla ^{p}\left( \left\langle \varphi ^{2}\right\rangle
-\left\langle \varphi ^{2}\right\rangle _{\mathrm{BD}}\right) &=&-\frac{D-1}{%
\alpha ^{2}}\left( \left\langle \varphi ^{2}\right\rangle -\left\langle
\varphi ^{2}\right\rangle _{\mathrm{BD}}\right) ,  \notag \\
\nabla _{1}\nabla _{1}\left( \left\langle \varphi ^{2}\right\rangle
-\left\langle \varphi ^{2}\right\rangle _{\mathrm{BD}}\right) &=&\left(
D-1\right) \cosh ^{2}(t/\alpha )\left( \left\langle \varphi
^{2}\right\rangle -\left\langle \varphi ^{2}\right\rangle _{\mathrm{BD}%
}\right) .  \label{Derphi}
\end{eqnarray}%
In the evaluation of the part in (\ref{TikVev}) containing the coincidence
limit we will consider the cases of even and odd $D$ separately. For even
values of $D$, the difference of the Hadamard functions is expressed as (\ref%
{Gdif2}). As before, putting $\theta =0$ and $\Delta \eta =0$, the term with
the coincidence limit is presented as%
\begin{eqnarray}
\lim_{x^{\prime }\rightarrow x}\partial _{1^{\prime }}\partial _{1}\left[
G\left( x,x^{\prime }\right) -G_{\mathrm{BD}}\left( x,x^{\prime }\right) %
\right] &=&-\frac{2(-2\pi )^{-D/2}}{\left[ \alpha \sinh (t/\alpha )\right]
^{D-1}}\int_{0}^{\infty }dz\,\frac{1}{e^{2\pi z}+1}  \notag \\
&&\times \lim_{x^{\prime }\rightarrow x}\partial _{\Delta r}^{2}\frac{%
P_{iz-1/2}^{D/2-1}\left( w\right) }{(w^{2}-1)^{(D-2)/4}}.  \label{lim1}
\end{eqnarray}%
By taking into account $\partial _{\Delta r}^{2}=(\sqrt{w^{2}-1}\partial
_{w})^{2}$ and by using the recurrence relations for the associated Legendre
function, it can be shown that%
\begin{equation}
\left( \sqrt{w^{2}-1}\frac{\partial }{\partial w}\right) ^{2}\frac{%
P_{iz-1/2}^{D/2-1}\left( w\right) }{(w^{2}-1)^{(D-2)/4}}=\frac{%
P_{iz-1/2}^{D/2+1}\left( w\right) }{(w^{2}-1)^{\left( D-2\right) /4}}+w\frac{%
P_{iz-1/2}^{D/2}\left( w\right) }{(w^{2}-1)^{D/4}}.  \label{RelP3}
\end{equation}%
From (\ref{P1}) it follows that the contribution of the first term in the
right-hand side of (\ref{RelP3}) vanishes in the limit $w\rightarrow 0$ and
one obtains%
\begin{eqnarray}
\lim_{x^{\prime }\rightarrow x}\partial _{1^{\prime }}\partial _{1}\left[
G\left( x,x^{\prime }\right) -G_{\mathrm{BD}}\left( x,x^{\prime }\right) %
\right] &=&-\frac{\left[ \alpha \sinh (t/\alpha )\right] ^{1-D}}{2^{D}\pi
^{D/2+1}\Gamma (D/2+1)}  \notag \\
&&\times \int_{0}^{\infty }dz\,e^{-\pi z}\left\vert \Gamma \left( \left(
D+1\right) /2+iz\right) \right\vert ^{2}.  \label{lim2}
\end{eqnarray}%
With this result, from (\ref{TikVev}) we find the expression for $\Delta
\left\langle T_{1}^{1}\right\rangle $. For the coefficient in (\ref{Tstruc})
one gets%
\begin{eqnarray}
C_{D} &=&-\frac{\pi ^{-D/2-1}}{2^{D}\Gamma (D/2)}\int_{0}^{\infty
}dz\,e^{-\pi z}z^{2}\left\vert \Gamma \left( \left( D-1\right) /2+iz\right)
\right\vert ^{2}  \notag \\
&=&-\frac{2^{1-D}\pi ^{-D/2}}{\Gamma (D/2)}\int_{0}^{\infty }dz\,\frac{%
z^{D}A_{D}(z)}{e^{2\pi z}+1},  \label{CDev}
\end{eqnarray}%
with $A_{D}$ defined in (\ref{AD}). In the special case $D=4$ we obtain%
\begin{equation}
C_{4}=-\frac{3}{2^{9}\pi ^{7}}\left[ \pi ^{2}\zeta (3)+15\zeta (5)\right] ,
\label{C4}
\end{equation}%
with $\zeta (x)$ being the Riemann zeta function.

For odd values of $D$ we can put $\theta =0$ and $\Delta \eta =0$ in (\ref%
{Gdif}). The limit is reduced to%
\begin{equation}
\lim_{x^{\prime }\rightarrow x}\partial _{1^{\prime }}\partial _{1}\left[
G_{0}\left( x,x^{\prime }\right) -G_{0}^{\mathrm{(BD)}}\left( x,x^{\prime
}\right) \right] =-\frac{(-2\pi )^{-\frac{D+1}{2}}}{\left[ \alpha \sinh
(t/\alpha )\right] ^{D-1}}\lim_{r^{\prime }\rightarrow r}\partial _{\Delta
r}^{2}\partial _{w}^{\frac{D-3}{2}}g(w).  \label{DGlim}
\end{equation}%
Combining this result with (\ref{Derphi}) and by using the expansion (\ref%
{gexp}) we can show that%
\begin{equation}
C_{D}=\frac{\left( D-1\right) ^{2}b_{D}-4Db_{D+2}}{48(2\pi )^{\frac{D+1}{2}}}%
.  \label{CDn}
\end{equation}%
In particular, one has%
\begin{equation}
C_{3}=-\frac{1}{480\pi ^{2}},\;C_{5}=-\frac{31}{60480\pi ^{3}}.  \label{C35}
\end{equation}%
The result for $D=3$ coincides with that found in \cite{Pfau82}. As seen, in
both cases of odd and even $D$ the energy density in the H-vacuum is smaller
than the one for the BD vacuum.

Similar to the case of the field squared, it can be shown that for odd $D$
the coefficient $C_{D}$ is presented in the integral form%
\begin{equation}
C_{D}=-\frac{2^{1-D}\pi ^{-D/2}}{\Gamma (D/2)}\int_{0}^{\infty }dz\,\frac{%
z^{D}A_{D}(z)}{e^{2\pi z}-1},  \label{CDodd}
\end{equation}%
where $A_{D}(z)$ is given by (\ref{AD}). Combining the results for even and
odd $D$, the coefficient in the expression (\ref{Tstruc}) for the VEV of the
energy-momentum tensor is written as
\begin{equation}
C_{D}=-\frac{2^{1-D}\pi ^{-D/2}}{\Gamma (D/2)}\int_{0}^{\infty }dz\,\frac{%
z^{D}A_{D}(z)}{e^{2\pi z}+(-1)^{D}}.  \label{CDevod}
\end{equation}

\subsection{Density of states and asymptotics}

Introducing the energy $E=z/\alpha $, the energy density for the H-vacuum is
written in the form%
\begin{equation}
\left\langle T_{0}^{0}\right\rangle =\left\langle T_{0}^{0}\right\rangle _{%
\mathrm{BD}}-\frac{\sinh ^{-D-1}(t/\alpha )}{2^{D-1}\pi ^{D/2}\Gamma (D/2)}%
\int_{0}^{\infty }dE\,\frac{E^{D}A_{D}(\alpha E)}{e^{2\pi \alpha E}+(-1)^{D}}%
.  \label{T00}
\end{equation}%
This shows the thermal nature of the BD vacuum with respect to the H-vacuum
with the temperature $T=1/(2\pi \alpha )$. Denoting by $\rho (E)dE$ the
number of states in the energy range $(E,E+dE)$, from (\ref{T00}) we read
the density of states%
\begin{equation}
\rho (E)=\frac{2E^{D-1}A_{D}(\alpha E)}{\left( 4\pi \right) ^{D/2}\Gamma
(D/2)}.  \label{roE}
\end{equation}%
The same expression is obtained when one considers the thermal properties of
the Minkowski vacuum with respect to the Fulling-Rindler vacuum in flat
spacetime. Note that the density of states $\rho _{\mathrm{M}}(E)$ for zero
spin massless particles in Minkowski spacetime is obtained by integrating
the number of states $d^{D}\mathbf{p}/(2\pi )^{D}$ over the angles
determining the direction of the momentum $\mathbf{p}$ and is related to (%
\ref{roE}) by the formula $\rho (E)=\rho _{\mathrm{M}}(E)A_{D}(\alpha E)$.
It is of interest to note that in even number of spatial dimensions the
average number of particles is given by Fermi-Dirac distribution. Similar
features for scalar Rindler particles in Minkowski vacuum and in the
response of particle detectors have been already discussed in the literature
\cite{Taga85,Oogu86,Taga86,Jenn10}.

As it has been mentioned above, the relation between the VEVs in the BD and
H-vacuum states is similar to that for the Minkowski and Fulling-Rindler
vacua in flat spacetime. This is related to the conformal connection between
the dS and Rindler spacetimes. To show that (see also \cite{Birr82} for the
case $D=3$) let us consider the coordinate transformation $\left( \tau _{%
\mathrm{R}},\chi ,\mathbf{x}_{\mathrm{R}}\right) \rightarrow (\eta
,r,\vartheta ,\phi )$ with
\begin{equation}
\tau _{\mathrm{R}}=\frac{\eta }{\alpha },\;\chi =\frac{\alpha }{\cosh
r-\sinh r\cos \theta _{1}},\;x_{\mathrm{R}}^{l}=\chi w^{l}\sinh
r,\;l=2,\ldots ,D.  \label{ct}
\end{equation}%
In these relations%
\begin{eqnarray}
w^{2} &=&\sin \theta _{1}\cos \theta _{2},\ldots ,\;w^{D-2}=\cos \theta
_{D-2}\prod_{i=1}^{D-3}\sin \theta _{i},  \notag \\
w^{D-1} &=&\cos \phi \prod_{i=1}^{D-2}\sin \theta _{i},\;w^{D}=\sin \phi
\prod_{i=1}^{D-2}\sin \theta _{i}.  \label{wD}
\end{eqnarray}%
The Rindler line element (\ref{ds2Ri}) is transformed to%
\begin{equation}
ds_{\mathrm{R}}^{2}=\chi ^{2}\left( d\eta ^{2}/\alpha ^{2}-dr^{2}-\sinh
^{2}rd\Omega _{D-1}^{2}\right) .  \label{ds2Ri2}
\end{equation}%
Comparing with (\ref{ds2c}), the conformal relation
\begin{equation}
ds^{2}=\Omega _{\mathrm{R}}^{2}ds_{\mathrm{R}}^{2},\;\Omega _{\mathrm{R}%
}^{2}=\frac{\left( \cosh r-\sinh r\cos \theta _{1}\right) ^{2}}{\sinh
^{2}\left( \eta /\alpha \right) },  \label{ds2R}
\end{equation}%
is seen between the dS and the Rindler spacetimes. The conformal counterpart
of the H-vacuum in dS spacetime is the Fulling-Rindler vacuum in flat
spacetime and this explains the above mentioned similarity of the
corresponding relations between the VEVs. The conformal relation between the
Rindler and dS spacetimes has been used in \cite{Saha04} for the
investigation of the vacuum average value of the energy-momentum tensor
induced by a curved brane in dS spacetime.

Now let us consider the flat spacetime limit of the results given above. In
that limit $\alpha \rightarrow \infty $ and the line element (\ref{dsH}) is
reduced to the one for the Milne universe
\begin{equation}
ds_{\mathrm{Milne}}^{2}=dt^{2}-t^{2}(dr^{2}+\sinh ^{2}rd\Omega _{D-1}^{2}).
\label{dsMilne}
\end{equation}%
The flat spacetime counterpart of the H-vacuum is the conformal vacuum in
the Milne universe (for the investigation of the corresponding VEVs induced
by a spherical boundary see \cite{Saha20}). The flat spacetime limit for the
BD vacuum corresponds to the Minkowski vacuum. Assuming that the VEVs for
the latter are renormalized to zero, for the VEVs in the conformal vacuum of
the Milne universe from (\ref{phi22}) we get
\begin{eqnarray}
\left\langle \varphi ^{2}\right\rangle _{\mathrm{Milne}} &=&-\frac{B_{D}}{%
t^{D-1}},  \notag \\
\left\langle T_{i}^{k}\right\rangle _{\mathrm{Milne}} &=&\frac{C_{D}}{t^{D+1}%
}\mathrm{diag}\left( 1,-1/D,\cdots ,-1/D\right) ,  \label{Milne}
\end{eqnarray}%
with the coefficients (\ref{BD}) and (\ref{CDevod}). These expressions give
the leading terms in the asymptotic expansions for the VEVs of the field
squared and energy-momentum tensor for a massive scalar field with general
curvature coupling parameter at early stages of the cosmological expansion ($%
t\rightarrow 0$). As seen, at early stages the VEVs are large and and the
backreaction of quantum effects is essential.

At late stages of the expansion, $t\gg \alpha $, the difference in the VEVs
for the H- and BD vacua is suppressed by the exponential factors $%
e^{-(D-1)t/\alpha }$ and $e^{-(D+1)t/\alpha }$ for the field squared and
energy-momentum tensor, respectively. This is an example of the general
result from \cite{Ande01}, stating that under the condition $m^{2}+\xi R>0$
the BD vacuum is the future attractor for cosmological solutions driven by a
scalar field with mass $m$ and curvature coupling parameter $\xi $ (for a
more recent discussion of the attractor properties of the BD state in
interacting field theories see \cite{Buch17}).

We have considered the VEV of the energy-momentum tensor in a fixed
background. Among the interesting topics of quantum field theory in curved
spacetime is the investigation of backreaction of quantum effects on the
classical geometry (see, for example, \cite{Birr82,Buch92}). That is done on
the base of semiclassical Einstein equations with the expectation value of
the energy-momentum tensor of quantum fields as an additional gravitational
source. In particular, motivated by stability issues of dS spacetime, the
backreaction of quantum fields on the properties of the BD state has been
widely considered in the literature by using variety of methods (see, e.g.,
\cite{Aals19,Mats19} and references cited there). The part in (\ref{Tstruc})
corresponding to the BD vacuum state renormalizes the bare cosmological
constant. From the point of view of backreaction, interesting effects come
from the contribution in (\ref{Tstruc}) corresponding to the difference in
the properties of the BD and H- vacua. Its role becomes essential at
early stages of the expansion and the backreaction effects should be taken
into account. The energy density for that part is negative and it violates
the energy conditions in general relativity. The corresponding equation of
state is of the radiation type ($p=\varepsilon /D$, with the pressure $p$
and the energy density $\varepsilon $), though with the negative energy
density. We expect that in a more general background, described by the line
element (\ref{dsH}) with a general scale factor $a(t)$ instead of $\alpha
\sinh \left( t/\alpha \right) $, the difference of the vacuum
energy-momentum tensors for a conformally coupled massless field in two
homogeneous vacuum states will have a structure similar to the last term in (%
\ref{Tstruc}) with the same replacement $\alpha \sinh \left( t/\alpha
\right) \rightarrow a(t)$. The corresponding cosmological dynamics can be
investigated based on the Friedmann equation for open cosmological models in
the presence of a positive cosmological constant. We plan to return to this
point in future work. 

The results obtained above can be applied to dS
bubbles with different vacua in the interior and exterior regions. For those
geometries the last term in the right-hand side of (\ref{Tstruc}) is present
in the region with the H-vacuum and is absent in the region with the BD
vacuum. Note that in those models additional contributions to the VEVs come
from the boundary, separating two phases. These are the Casimir type
contribution to the vacuum characteristics (see \cite{Bell14} for a general
discussion).

\section{Conclusion}

\label{sec:Conc}

We have investigated the mean field squared and the VEV of the
energy-momentum tensor for a scalar field prepared in the H-vacuum of dS
spacetime. Bearing in mind possible applications in field-theoretical models
with extra spatial dimensions, a general number of spacetime dimension is
considered. The properties of the vacuum state are described by two-point
functions and as the first step the Hadamard function is discussed. The
corresponding expression is decomposed into two contributions. The first one
presents the Hadamard function for the BD vacuum and the second one
describes the difference in the correlations of the vacuum fluctuations in
those two vacua. With that representation, the renormalization of the VEVs
for the H-vacuum is reduced to the renormaliztion for the BD vacuum state.
The latter is well investigated in the literature. The expression for the
Hadamard function is essentially simplified in the special case of a
conformally coupled massless scalar field. The corresponding expressions for
odd and even values of the spatial dimension are given by (\ref{GconfO}) and
(\ref{GconfEv}), respectively. The Hadamard function for the BD vacuum is
expressed as (\ref{GBDconf3}) and (\ref{Rel2}).

Given the Hadamard functions, the differences in the mean field squared and
the VEV of the energy-momentum tensor for the H- and BD vacua are obtained
by making use of the formulas (\ref{phi2}) and (\ref{TikVev}). The
divergences in the coincidence limit are the same for those states and the
renormalization is not required for the differences. The corresponding
expressions for general case of spatial dimension are given by (\ref{phi22})
and (\ref{Tstruc}). The coefficients in those formulas are expressed as (\ref%
{BD}) and (\ref{CDevod}). These expressions prove the thermal nature of the
BD state with respect to the H-vacuum. The corresponding density of states
is expressed as (\ref{roE}) with the factor $A_{D}(\alpha E)$ defined by (%
\ref{AD}). The latter is interpreted in terms of the ratio of the densities
of states for a conformally coupled massless field in dS spacetime and for
massless zero spin particles in Minkowski spacetime. It is of interest to
note that in odd number of spacetime dimensions the thermal distribution is
of Fermi-Dirac type. The essential difference of the dynamics of quantum
fields in odd and even dimensional dS spacetimes is also seen in the
analysis of the particle production process. As it has been discussed in
\cite{Bous02}, there is no particle production in odd dimensions and
particles are created in even number of dimensions. At late stages of the dS
expansion the difference between the VEVs in the H- and BD vacua is
exponentially suppressed and this is in agreement with the result on the BD
vacuum as the future attractor for a general class of cosmological
solutions. At early stages the contributions corresponding to the BD vacuum
are subdominant and the behavior of the VEVs is essentially different. At
those stages the vacuum energy-momentum tensor is large and the backreaction
of quantum effects on the spacetime geometry should be taken into account.

\section*{Acknowledgments}

A.A.S. was supported by the grant No. 20RF-059 of the Committee of Science
of the Ministry of Education, Science, Culture and Sport RA. T.A.P. was
supported by the Committee of Science of the Ministry of Education, Science,
Culture and Sport RA in the frames of the research project No. 20AA-1C005.
The work was partly supported by the "Faculty Research Funding Program" (PMI
Science and Enterprise Incubator Foundation).

\end{document}